\documentclass[aps, prl, twocolumn, superscriptaddress, nofootinbib, showkeys, floatfix]{revtex4-1}
\pdfoutput=1
\usepackage{graphicx}   % need for figures
\usepackage{amssymb,amsmath}

\usepackage{natbib,hyperref}
\usepackage{hyperref}
\usepackage{color}
\usepackage{pdfpages}

\makeatletter
\AtBeginDocument{\let\LS@rot\@undefined}
\makeatother

\begin{document}

%%%%%%%%%%%%%%%%%%%%%%%%%%%%%%%%%%%%%%%%%%%%%%%%%%%%%
%%
%%						Title
%%
%%
\title{Engineering Multiple Topological Phases in Nanoscale\\ Van der Waals Heterostructures: Realisation of $\alpha$-Antimonene}

\author{T. M\"arkl}
\affiliation{The MacDiarmid Institute for Advanced Materials and Nanotechnology, Department of Physics and Astronomy, University of Canterbury, Private  Bag 4800, Christchurch 8140, New Zealand}
\author{P. J. Kowalczyk}
\affiliation{Department of Solid State Physics, Faculty of Physics and Applied Informatics, University of Lodz, 90-236 Lodz, Pomorska 149/153, Poland}
\author{M. Le Ster}
\affiliation{The MacDiarmid Institute for Advanced Materials and Nanotechnology, Department of Physics and Astronomy, University of Canterbury, Private  Bag 4800, Christchurch 8140, New Zealand}
\author{I. V. Mahajan}
\affiliation{The MacDiarmid Institute for Advanced Materials and Nanotechnology, Department of Physics and Astronomy, University of Canterbury, Private  Bag 4800, Christchurch 8140, New Zealand}
\author{H. Pirie}
\affiliation{The MacDiarmid Institute for Advanced Materials and Nanotechnology, Department of Physics and Astronomy, University of Canterbury, Private  Bag 4800, Christchurch 8140, New Zealand}
\author{Z. Ahmed}
\affiliation{The MacDiarmid Institute for Advanced Materials and Nanotechnology, Department of Physics and Astronomy, University of Canterbury, Private  Bag 4800, Christchurch 8140, New Zealand}
\author{G. Bian}
\affiliation{Department of Physics and Astronomy, University of Missouri, Columbia, Missouri 65211, USA}
\author{X. Wang}
\affiliation{College of Science, Nanjing University of Science and Technology, Nanjing 210094, China}
\affiliation{Department of Physics, University of Illinois at Urbana-Champaign, 1110 West Green Street, Urbana, Illinois 61801-3080, USA}
\author{T.-C. Chiang}
\affiliation{Department of Physics, University of Illinois at Urbana-Champaign, 1110 West Green Street, Urbana, Illinois 61801-3080, USA}
\author{S. A. Brown}
\email[Corresponding author: ]{simon.brown@canterbury.ac.nz}
\affiliation{The MacDiarmid Institute for Advanced Materials and Nanotechnology, Department of Physics and Astronomy, University of Canterbury, Private Bag 4800, Christchurch 8140, New Zealand}


\date{\today}

%%%%%%%%%%%%%%%%%%%%%%%%%%%%%%%%%%%%%%%%%%%%%%%%%%%%%
%%
%%						abstract
%%
%%
\begin{abstract}
\noindent Van der Waals heterostructures have recently been identified as providing many opportunities to create new two-dimensional materials, and in particular to produce materials with topologically-interesting states. Here we show that it is possible to create such heterostructures with multiple topological phases in a single nanoscale island. We discuss their growth within the framework of diffusion-limited aggregation, the formation of moir\'e patterns due to the differing crystallographies of the materials comprising the heterostructure, and the potential to engineer both the electronic structure as well as local variations of topological order. In particular we show that it is possible to build islands which include both the hexagonal $\beta$- and rectangular $\alpha$-forms of antimonene, on top of the topological insulator $\alpha$-bismuthene. This is the first experimental realisation of $\alpha$-antimonene, and we show that it is a topologically non-trivial material in the quantum spin Hall class.
\end{abstract}

\maketitle

%%%%
\section*{Introduction}
%%%%
Van der Waals heterostructures consist of multilayers of weakly-coupled few-monolayer-thick sheets of materials \cite{Hunt2013, Geim-N499-419-13, Novoselov2016, Ajayan-PT69-38-16}. This weak coupling means that the intrinsic properties of individual layers of the material of interest may be preserved, only mildly perturbed by interactions with other layers, and so provides potential for realisation of new topologically interesting materials~\cite{Ajayan-PT69-38-16}. Topological insulators are materials in which strong spin-orbit coupling (SOC) leads to an inversion of the usual band structure, and hence to unusual topologically protected boundary states~\cite{Hasan-RoMP82-3045-10, Qi-RoMP83-1057-11}. In the two-dimensional (2D) systems of interest here, the topologically protected states are electronic edge states in which spin and momentum are locked together so that backscattering is forbidden. The dissipationless 1D spin currents in these systems
provide a platform for the fabrication of spin-selective electronic devices~\cite{Vandenberghe2017, Tong-NP--16}.

To date, only a very limited range of van der Waals heterostructures have been reported, and the materials that have been investigated have almost exclusively hexagonal symmetry~\cite{Mannix-NRC1-14-17, Hunt2013,Novoselov2016, Geim-N499-419-13, Lee-NN9-676-14, Ferrari-N7-4598(15)b}. While such graphene-like~\cite{Novoselov-S306-666-04} structures are very promising, an ability to engineer heterostructures with different crystal symmetries could open up significant new opportunities in both fundamental physics and electronic device fabrication. Moreover, incommensurate moir\'e structures, which modulate the electronic states in 2D materials, can lead to the emergence of a fractal spectrum of electronic states containing new Dirac points~\cite{Ponomarenko-N--13, Dean}, and could provide a platform for the observation of predicted spatial modulation of local topological phases~\cite{Tong-NP--16}.

\begin{figure*} 
\begin{center}
\includegraphics[width=153mm]{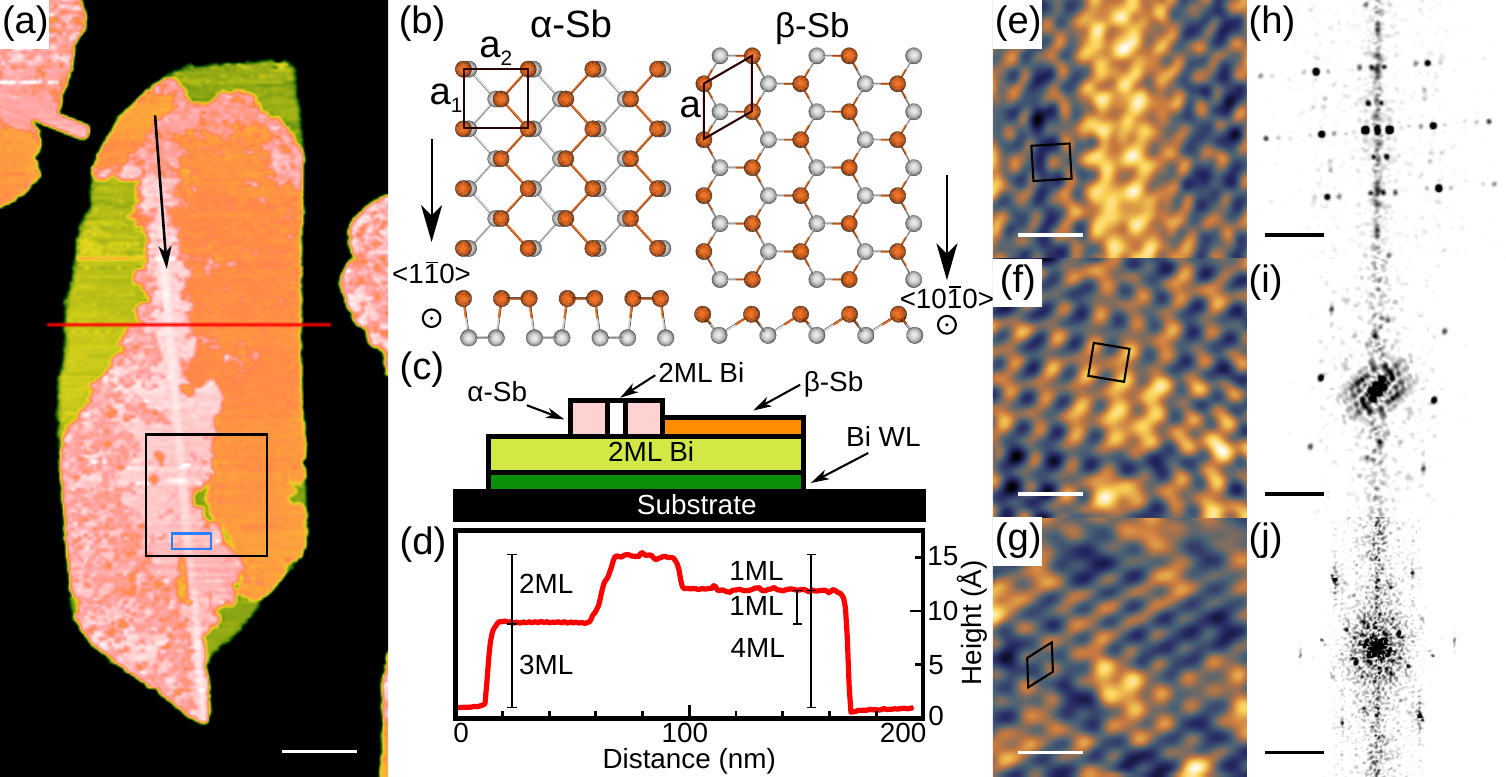}
\caption{
\label{fig:topo} Morphology and atomic structure of the layered nanoislands. \textbf{(a)} Large-scale STM image of a typical van der Waals heterostructure comprising the $\alpha$-Bi island base (light green), 2\,ML-$\alpha$-Sb (pink) and 1\,ML-$\beta$-Sb (orange). (Imaging conditions: +400\,mV, 10\,pA, 295\,K). The scale bar corresponds to 50\,nm. The arrow indicates the location of an additional 2\,ML thick Bi stripe on top of the base. The regions marked by rectangles are shown in detail in Figure~\ref{fig:moire}(b) (black rectangle) and Supplementary Figure~S1(d) (blue rectangle). \textbf{(b)} Schematic top and side views (created with VESTA \cite{VESTA}) of the crystal structures of single sheets of $\alpha$-Sb and $\beta$-Sb, respectively; the unit cells are marked in black. \textbf{(c)} Schematic side view of the island showing the stacking of the different layers, with colors corresponding to the topography image in (a). The wetting layer (Bi WL)~\cite{Kowalczyk-APL100-151904(12)} is shown separately in dark green. \textbf{(d)} Height profile at the indicated position (red line in (a)). \textbf{(e)-(g)} Atomically resolved STM images of $\alpha$-Bi, 2\,ML-$\alpha$-Sb, and 1\,ML-$\beta$-Sb, respectively (scale bar: 1\,nm; imaging conditions: 50\,mV, 200\,pA for (e); 100\,mV, 50\,pA for (f) and (g); all at 295\,K.) Larger scale atomically-resolved images are presented in Supplementary Figure~S1. \textbf{(h)-(j)} show the inner regions of the 2D Fourier transforms of the respective raw images. (Scale bar: 1.3\,\AA$^{-1}$)
}
\end{center}
\end{figure*}

Bismuth is a key ingredient in many topologically interesting materials because of its strong SOC. As a thin film, Bi exists in two forms: the [Black Phosphorous (BP)-like, rectangular symmetry] $\alpha$-form and the [hexagonal (111)-oriented] $\beta$-form which are both called 'bismuthene'. It has been shown recently that $\alpha$-bismuthene~\cite{Lu-NL-15-80(15)b}, $\beta$-bismuthene~\cite{Wada-PRB83-121310(11)} and even bulk Bi~\cite{Ito-PRL117-236402-16} are topological insulators. Antimony, which sits above Bi in group 15 of the periodic table, has similar allotropes but is much less well explored. It is known that multilayer $\beta$-antimonene is a topological insulator, but that coupling between the surfaces destroys the topological states in few-monolayer-thick films~\cite{Bian-PRL107-36802-11, Zhang2012}, unless sufficient strain is applied~\cite{Zhao2015}. In contrast, despite significant theoretical interest~\cite{Wang-AAMI7-11490-15, Pumera-ADMA-2017, Aktuerk-PRB91-235446-15, Ji-NC7-13352-16, Lu-NCM2-16011-16}, $\alpha$-antimonene has never been realized experimentally and its topological properties have not been investigated. The $\alpha$-phases of Sb and Bi are particularly interesting because they have structures that are homologous to those recently used to build novel BP transistors~\cite{Li-NN9-372-14}. If similar transistors could be fabricated with built-in strong SOC their topologically-protected edge states could be used in spin-selective devices allowing manipulation and robust transport of quantum information -- the present antimonene / bismuthene structures, grown on microelectronics-compatible substrates, are a significant step towards this goal.

We report the realisation of van der Waals heterostructures comprising multiple bismuthene and antimonene allotropes. These include: (i) $\alpha$-bismuthene, (ii) one monolayer (ML)\footnote{We define~\cite{Scott-PRB72-205423(05)} a monolayer to be a single sheet of atoms which in all cases is 3\,\AA{} to 4\,\AA{} thick; these monolayers are sometimes called a 'bilayer' in the literature because of puckering which results in atoms that lie in two separate planes (see Figure~\ref{fig:topo}(b)). We use the suffix '-ene' to denote the thinnest possible films for each structure: 1\,ML-$\beta$-Sb, 2\,ML-$\alpha$-Sb, and 2\,ML-$\alpha$-Bi.} thick sheets of $\beta$-antimonene (1\,ML-$\beta$-Sb) and (iii) 2\,ML thick sheets of the never previously experimentally observed $\alpha$-antimonene (2\,ML-$\alpha$-Sb). We investigate these structures using scanning tunnelling microscopy (STM), scanning tunnelling spectroscopy (STS), and density functional theory to show that the new allotrope 2\,ML-$\alpha$-Sb is topologically non-trivial and energetically preferred over 2\,ML-$\beta$-Sb because of interactions with the underlying Bi islands. All three allotropes exhibit distinctive moir\'e patterns that provide potential for engineering spatially dependent topological order~\cite{Tong-NP--16}.

\section*{Methods}
In our experiments Bi nanoislands (which have been intensively studied previously~\cite{Scott-PRB73-205424(06), Kowalczyk-SS605-659(11), Kowalczyk-APL100-151904(12), Kowalczyk-NL13-43(13), Nagao-PRL93-105501(04), Zhang-ACIE55-1666(16)}) serve as a basis for the growth of the antimonenes. For growing those Bi islands we have used MoS${_2}$ substrates as well as the more usual graphite (HOPG) and obtain qualitatively similar results in both cases, but focus here on the MoS${_2}$ samples for three different reasons: (i) MoS${_2}$ is a semiconductor which provides opportunities for fabrication of electronic devices, (ii) the difference in the interactions between the Bi and the MoS${_2}$ (compared to HOPG) lead to subtle but interesting structural differences, and (iii) neither antimonene nor bismuthene have been reported on MoS${_2}$ before. Sample preparation was carried out in-situ under UHV conditions; the nanostructures presented here were grown by first thermally evaporating 2\,ML of Bi onto MoS${_2}$ substrates (held at 295\,K), followed by 0.6\,ML of Sb. STM/STS measurements were performed either at room temperature or $\sim$50\,K using a commercial Omicron VT AFM/STM with Pt/Ir tips. Imaging and STS conditions are stated in the figure captions. The lateral resolution of the STM is calibrated using atomically resolved images of MoS$_2$.

The first-principles calculations are similar to those used to predict the band structure of $\alpha$-Bi~\cite{Kowalczyk-NL13-43(13), Bian-PRB80-245407(09)}. They were performed with ABINIT package within plane-wave expansion and pseudopotential framework. In order to take relativistic effects into account, the Hartwigsen-Goedecker-Hutter (HGH) pseudopotential was used in the calculations. The cutoff of electron kinetic energy was set to be 400\,eV. The k-sampling was carried out on a $\Gamma$-centered 13$\times$13$\times$1 grid based on Monkhorst-Pack method. The films were modelled by a periodic array of slabs separated by a 15\,\AA{} vacuum gap between neighboring layers. The atomic positions, including the in-plane lattice constants, were allowed to relax by energy minimization until the Hellmann-Feynman forces were reduced to below 1.0$\times$×10$^{-5}$\,eV/\AA.

\section*{Results and Discussion}

Figure~\ref{fig:topo}(a) shows a typical STM image of a nanoscale Bi-island decorated with antimonene. The (110)-oriented Bi-island is 3\,ML thick\footnote{Note that a wetting layer exists underneath the 2\,ML-$\alpha$-Bi. \cite{Kowalczyk-SS605-659(11), Kowalczyk-NL13-43(13)}} and exhibits three atomically flat regions of different height (line profile in Figure~\ref{fig:topo}(d)): the 3\,ML bismuth island base (green), a 5\,ML high region (pink), and a region which is 1\,ML higher than the base (orange). On careful inspection the 5\,ML high central region comprises a Bi stripe (arrowed) which is absolutely linear, with straight edges running parallel to the Bi $\langle1\bar{1}0\rangle$ direction~\cite{Scott-PRB73-205424(06), Scott-TEPJD-AMOaPP39-433-06, Scott-PRB72-205423(05)} surrounded by a laterally less well-defined Sb region of variable width. This Sb film is 2\,ML high, and is supported by the 3\,ML Bi island base. As described in detail below, based on atomic resolution images and spectroscopic data, the differently coloured regions can be identified as 2\,ML-$\alpha$-Bi (green), 2\,ML-$\alpha$-Sb (pink), and 1\,ML-$\beta$-Sb (orange). Hence, in this one island, we see a heterostructure comprising three distinct topologically interesting materials (as shown schematically in Figure~\ref{fig:topo}(c)). 

\begin{figure}
\begin{center}
\includegraphics[width=74.5mm]{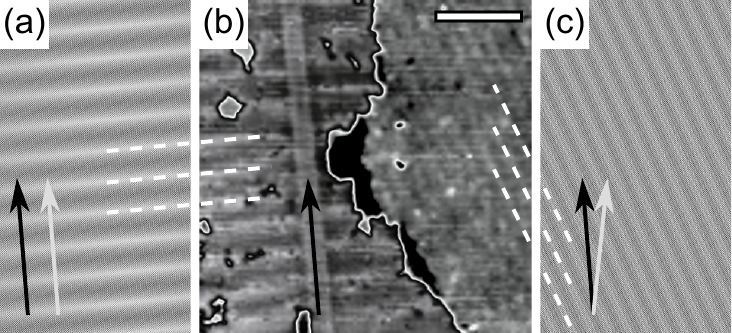}
\caption{
\label{fig:moire} Example of observed and simulated moir\'e patterns of 2\,ML-$\alpha$-Sb and 1\,ML-$\beta$-Sb on a 3\,ML Bi island. In all three panels the black arrow represents the Bi $\langle 1 \bar{1} 0 \rangle$ direction. Dashed white lines serve as guides to the eye. \textbf{(a)} Simulated moir\'e pattern of 2\,ML-$\alpha$-Sb. The grey arrow indicates the $\alpha$-Sb $\langle 1 \bar{1} 0 \rangle$ direction which is parallel to the Bi $\langle 1 \bar{1} 0 \rangle$ direction in this case. \textbf{(b)} STM image (200\,mV, 10\,pA, scale bar: 20\,nm) from inside the black rectangle in Figure~\ref{fig:topo}(a), enhanced to show both moir\'e patterns of $\alpha$-Sb and $\beta$-Sb. (Supplementary Figure~S2 shows an example in which different moir\'e patterns are observed due to differing orientations of the overlayer and underlayer.) \textbf{(c)} Simulated moir\'e pattern of 1\,ML-$\beta$-Sb. The grey arrow represents the $\beta$-Sb $\langle 1 0 \bar{1} 0 \rangle$ direction, which is rotated by 10.5$^\circ$ with respect to Bi $\langle 1 \bar{1} 0 \rangle$.}
\end{center}
\end{figure}

Figures~\ref{fig:topo}(e)-(g) show magnified atomic resolution images obtained from examples of each of the three phases of interest (see also Supplementary Figure S1). In each case the symmetry of the lattice is indicated, alongside fast Fourier transforms (Figures~\ref{fig:topo}(h)-(j)) of the images which allow determination of the lattice constants of the surface unit cells. We obtain $a^{\beta\text{-Sb}}\!=\!404$\,pm, $a_1^{\alpha\text{-Sb}}\!\!\times\!a_2^{\alpha\text{-Sb}}\!=\!429$\,pm$\times486$\,pm, and $a_1^{\alpha\text{-Bi}}\!\times\!\!a_2^{\alpha\text{-Bi}}\!=\!453$\,pm$\times487$\,pm. These values agree well with the result of our calculations and existing literature values \cite{Aktuerk-PRB91-235446-15, Zhao2015, Wu2016, Gibaja2016, Wang-AAMI7-11490(15), Lu-NL-15-80(15)b, Kowalczyk-PRB91-45434(15), Zhang-ACIE55-1666(16)}. Refer to Supplementary Table~SI for full details, including experimental uncertainties.

In our topography images we observe apparent spatial modulations of the height $\leq50$\,pm that are hardly visible in Figure~\ref{fig:topo}(a); hence, we show them on expanded scales in Figure~\ref{fig:moire} and Supplementary Figures~S1 and S2. In Figure~\ref{fig:moire}~(b) the $\alpha$-Sb fringes have a period of 7.3\,nm\,$\pm$\,0.6\,nm and are perpendicular to the Bi $\langle 1 \bar{1} 0\rangle$ direction; the $\beta$-Sb fringes are inclined with respect to the Bi $\langle 1 \bar{1} 0\rangle$ direction at an angle of $20^\circ\pm5^\circ$ and exhibit a period of 4.0\,nm$\pm$0.4\,nm.

To show that these fringes are moir\'e patterns we have carefully modelled them by creating superpositions of the two lattices with the software VESTA \cite{VESTA} and by varying the lattice constants and relative angles. Supplementary Figure~S3 exemplifies that the observed period and angle of the moir\'e pattern are very sensitive to these parameters.

We begin by considering the pattern that is observed on the 3\,ML-Bi bases due to the mismatch between the rectangular bismuth crystal structure and the underlying hexagonal MoS${_2}$. We find that the best agreement with the experimentally observed moir\'e pattern (Figure~S1~(a)) is achieved for Bi lattice constants that match exactly with those obtained from atomic resolution (453\,pm$\times$487\,pm). Hence, we use those lattice parameters to subsequently model the antimonene structures on $\alpha$-Bi.

The moir\'e fringes for 2\,ML-$\alpha$-Sb on $\alpha$-Bi are perpendicular to the $\alpha$-Bi stripe, consistent with atomic resolution results which show that $\langle1\bar{1}0\rangle_{Sb}$ is parallel to $\langle1\bar{1}0\rangle_{Bi}$ (Supplementary Figure~S1(d)). This means that $a_2^{Sb}\approx a_2^{Bi}=487$\,pm~\footnote{If there is an additional modulation along the length of the fringes its period must be at least twice the observed $\alpha$-Sb film width of $\sim$40\,nm in Fig.~\ref{fig:topo}. The corresponding lattice mismatch can be estimated to be $\leq$3\,pm.}, and only $a_1^{Sb}$ is to be determined. We find an optimal match (Fig. \ref{fig:moire}(a)) to the observed moir\'e pattern in Figure~\ref{fig:moire}~(b) using $a_1^{Sb}=426$\,pm, in good agreement with the atomic resolution value.

In the case of 1\,ML-$\beta$-Sb on $\alpha$-Bi, we find the best match (Fig. \ref{fig:moire}(c)) for the moir\'e pattern in Figure~\ref{fig:moire}~(b) by using a lattice constant of 413\,pm. This is slightly larger than the result from atomic resolution and calculations but still in reasonable agreement. Supplementary Figure~S4 shows the modelled moir\'e patterns for both $\alpha$-Sb and $\beta$-Sb in more detail.

We now consider the growth mechanism of these van der Waals heterostructures. It is well-established that both Sb and Bi nanostructures grow by diffusion and aggregation on atomically-flat substrates like HOPG~\cite{Scott-PRB73-205424(06), Kowalczyk-NL13-43(13), Scott-TEPJD-AMOaPP39-433-06}. The present bismuth structures are very similar to those grown on HOPG but subtle differences include predominance of narrower 5\,ML stripes and broader 3\,ML island bases, and a shorter wavelength moir\'e pattern (Supplementary Figure~S1(a)) -- these differences originate from the different MoS${_2}$ and HOPG lattice constants.

The growth of the antimony layers on top of the bismuth structure is rather remarkable. It is known~\cite{Scott-TEPJD-AMOaPP39-433-06} that Sb prefers to form three-dimensional structures in comparison to the planar 2D structures preferred by Bi, and this is manifested in the present islands by up-diffusion of Sb from the MoS${_2}$ substrate onto the Bi island bases, no Sb is observed on the MoS$_2$. The amount of Sb that is directly deposited onto the top surface of the Bi islands is very small, and so the amount of Sb observed on top of the Bi island reflects the size of the capture zone on the MoS${_2}$ substrate (e.g. the bottom left of the island shown in Figure~\ref{fig:topo}(a) is covered with a thicker layer of Sb because of a large open MoS${_2}$ terrace on that side of the island). 

\begin{figure}%[h!]
\begin{center}
\includegraphics[width=74.5mm]{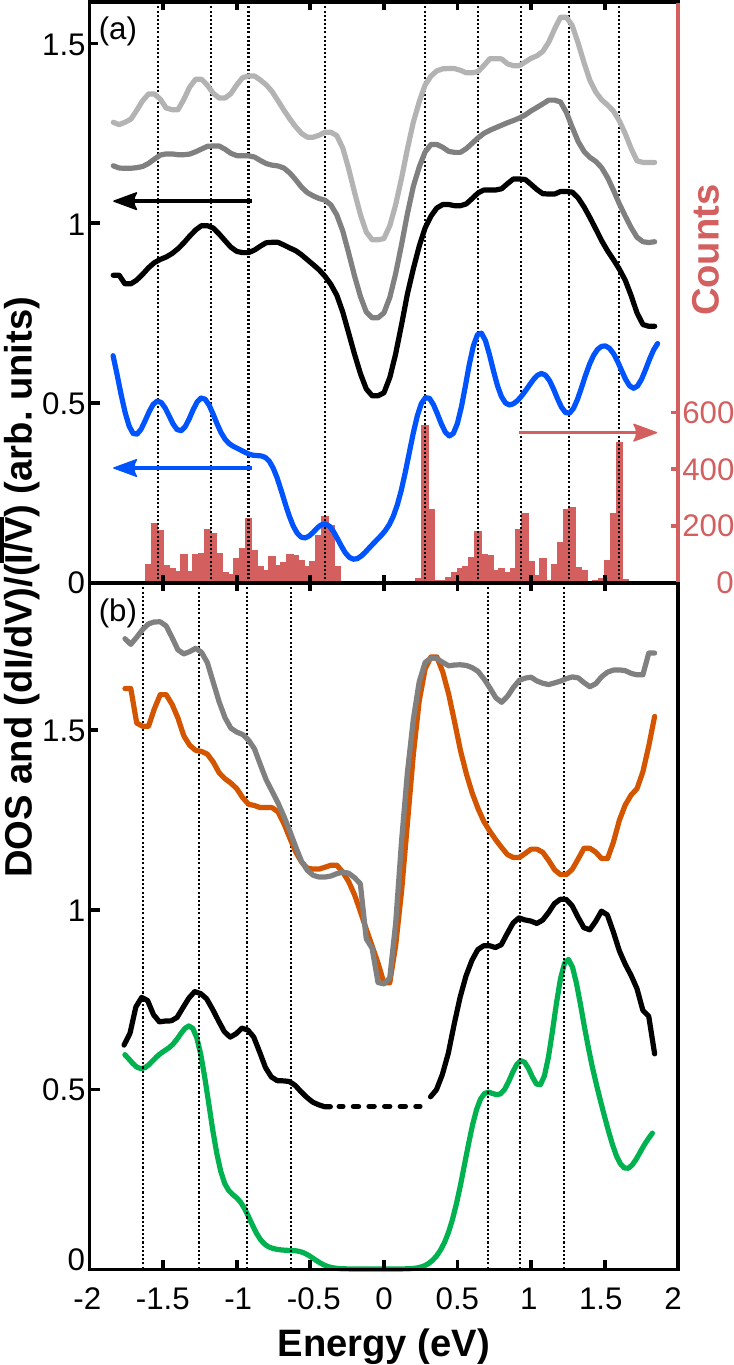}
\caption{
\label{fig:dos_spec} Comparison of experimental STS data and calculated DOS for relaxed antimonene structures. Dotted lines indicate the position of experimental peaks and serve as guides to the eye. \textbf{(a)}: Three examples of 2\,ML-$\alpha$-Sb spectra (black, dark and light grey; vertically offset for clarity) and DOS (blue). A histogram at the bottom of the panel shows the detected position of local maxima and shoulders in 850 individual spectra. The calculation was shifted by 140\,mV to lower energies giving a good agreement with the main peak positions of the histogram. \textbf{(b)} Comparison of calculated DOS (green) with a representative experimental spectrum for 1\,ML-$\beta$-Sb (black). This spectrum was obtained by subtracting the scaled spectrum from the 3\,ML-Bi island base (brown, shifted by 60\,mV to higher energies) from that of the original 1\,ML-$\beta$-Sb spectrum (grey). This subtraction procedure removes the contribution of the underlying 3\,ML-Bi layer, which dominates the measured data for 1\,ML-$\beta$-Sb because of the low density of states of the 1\,ML-$\beta$-Sb near zero bias. Noise in the central part of the black curve (due to the applied subtraction procedure) has been removed. STS spectra were obtained by numerical differentiation of recorded I-V curves ($\pm2$\,V, current set points ranging between 0.2\,nA and 1\,nA) after median and average filtering per curve, followed by normalization to $\overline{I/V}$ and spatial averaging of 150-200 resulting individual curves.}

\end{center}
\end{figure}

In the vast majority of islands the Bi base was covered by either antimonene or an additional 2\,ML-Bi stripe. Bare 3\,ML Bi regions were scarce and hence we conclude that on the Bi bases the antimony prefers to form a complete layer first before growing thicker layers. In particular, the 1\,ML-$\beta$-Sb regions often covered the whole width of an island whereas 2\,ML-$\alpha$-Sb was mostly observed adjacent to the additional Bi stripe. After the formation of a 1\,ML-$\beta$-Sb layer the arrival of further Sb leads to a restructuring into 2\,ML-$\alpha$-Sb. The necessary nucleation can occur directly on the 1\,ML-$\beta$-Sb but is greatly facilitated by the presence of the Bi stripes. According to this scenario, we believe that the disordered regions on the left of Figure~\ref{fig:topo}(a) are yet to be fully-crystallized as $\alpha$-Sb. Additional data (not shown) shows that these disordered regions re-order into the $\alpha$-phase over a period of weeks because they were only found close to the $\alpha$-Sb and only within a few days after deposition.

Thicker layers of $\beta$-Sb, such as 2\,ML-$\beta$-Sb, were never observed in our experiments, which strongly indicates that 2\,ML-$\alpha$-Sb is energetically preferred. Our first principles calculations find that the energies of free-standing slabs of these two structures are identical within the numerical accuracy of the simulations. The preference for the $\alpha$-phase in 2\,ML thick films can therefore be attributed to the weak interaction with the underlying Bi which has rectangular symmetry. Interestingly it is known that Bi grows initially in the $\alpha$-phase but re-orients into the $\beta$-phase at a critical thickness that depends on the substrate~\cite{Nagao-PRL93-105501(04),Scott-SS587-175-05}. Sb also grows in the $\beta$-phase for thicknesses greater than $\sim$20\,ML~\cite{Bian-PRL107-36802-11} which suggests that the $\alpha$-phase is preferred only for an intermediate thickness range. Deposition of additional Sb will likely lead to observation of $\alpha$-Sb in the sequence of thicknesses 4\,ML, 6\,ML, 8\,ML etc (as for Bi)~\cite{Nagao-PRL93-105501(04), Scott-PRB73-205424(06), Kowalczyk-SS605-659(11), Kowalczyk-APL100-151904(12), Kowalczyk-NL13-43(13), Zhang-ACIE55-1666(16)}, but it will be interesting to see at what thickness the transition to the $\beta$-phase occurs.

We have performed first principles calculations of the electronic states for both the experimentally observed 2\,ML-$\alpha$-Sb and 1\,ML-$\beta$-Sb structures (Supplementary Figure~S5). The band structure for 1\,ML-$\beta$-Sb is similar to that reported previously in Refs.~\cite{Wang-AAMI7-11490-15, Pumera-ADMA-2017, Aktuerk-PRB91-235446-15, Ji-NC7-13352-16}, and it is known to be topologically trivial~\cite{Bian-PRL107-36802-11, Zhang2012} (although under sufficient strain it becomes non-trivial~\cite{Zhao2015}). We now focus on the 2\,ML-$\alpha$-Sb films since their topological properties have never been reported previously. Our calculations show that there exists an indirect negative band gap separating the conduction and valence bands which allows the $\mathbb{Z}_2$ topological invariant to be defined: the crystal structure is centrosymmetric and thus the wavefunctions at each time reversal invariant momentum point carry definite parity eigenvalues. By using the Fu-Kane formula~\cite{Fu-PRB76-45302-07} the $\mathbb{Z}_2$ invariant can be straightforwardly calculated from the parity data in Supplementary Table~SII. Our results show that 2\,ML-$\alpha$-Sb films are topologically non-trivial, belonging to the quantum spin Hall class.

In Figure~\ref{fig:dos_spec}, we compare the calculated and measured density of states (DOS). Figure~\ref{fig:dos_spec}(a) shows experimental STS spectra (black, grey) for 2\,ML-$\alpha$-Sb which have deep minima at zero bias and weak peaks whose positions vary by $\sim\pm0.1$\,V (see the histogram). There is generally good agreement between positions of the peaks in the histogram and the peaks in the calculated DOS (which is shifted by 140\,mV, to account for charge transfer from the substrate~\cite{Kowalczyk-NL13-43(13)}) demonstrating that the calculated electronic structure matches that obtained experimentally. 

Figure~\ref{fig:dos_spec}(b) shows the measured STS data for 1\,ML-$\beta$-Sb (grey curve). At first sight, there is no agreement between the experimental data and the calculated DOS (green curve). However we find that the spectrum for 1\,ML-$\beta$-Sb is strikingly similar to that for the 3\,ML-Bi island base (brown curve) in the bias range $\pm0.5$\,V. As it is already known that $E_F$ lies in a band gap in 1\,ML-$\beta$-Sb, we assign the features in our spectra to the underlying Bi layers. And indeed, when the measured DOS for 3\,ML Bi is subtracted from that for the 1\,ML-$\beta$-Sb the result (black) is in very good agreement with the calculated DOS. 

The agreement between the calculated and measured DOS for both 1\,ML-$\beta$-Sb and 2\,ML-$\alpha$-Sb confirms that neither phase is significantly perturbed by interactions with the underlying Bi, and that the structures shown in Figure~\ref{fig:topo}(a) are indeed van der Waals heterostructures.

\section*{Summary}
One of the merits of our deposition technique is that it allows straightforward fabrication of heterostructures comprising materials with different thicknesses and different lateral dimensions~\cite{Scott-PRB73-205424(06), Scott-TEPJD-AMOaPP39-433-06}, and exhibiting a variety of moir\'e patterns. The combined influence of the moir\'e periodicity~\cite{Tong-NP--16}, inter-layer coupling, and strain on each of the materials within the heterostructures could allow exquisite control of the topology of their electronic states, and -- as demonstrated in the present work -- the realisation of new topological phases such as alpha-antimonene.

Finally, we note that the $\alpha$-phases of Bi and Sb have a BP-like structure which has not previously been employed in van der Waals heterostructures, and in which the absence of surface dangling bonds means the structure is resistant to oxidation~\cite{Kowalczyk-APL100-151904(12)}. Together with the semiconducting nature of the MoS${_2}$ substrate, and potential to fabricate similar structures on a range of other substrates (especially those with rectangular symmetry), these features open up many new possibilities for the engineering of spintronic / topological devices.

\section*{Notes}

The authors declare that they have no competing financial interests.

\section*{Acknowledgement}

This work was supported by the MacDiarmid Institute for Advanced Materials and Nanotechnology (T.M., M.L.S., I.V.M., Z.A., and S.A.B.) and the Marsden Fund (UOC1503, T.M., M.L.S., I.V.M., Z.A., and S.A.B.), the National Science Centre, Poland (DEC-2015/17/B/ST3/02362, P.J.K), the National Natural Science Foundation of China (11204133, X.X.W) and the U.S. National Science Foundation (NSF-DMR-1305583, T.-C.C.).

%%%%%%%%%
%%	References
%%

%%%%%%%%%%%

\bibliography{pub_sb_v02}

\clearpage



\section*{Supplementary material}

# Supplementary Information − Engineering Multiple Topological Phases in Nanoscale Van der Waals Heterostructures: Realisation of α-Antimonene

T. Märkl,[1] P. J. Kowalczyk,[2] M. Le Ster,[1] I. V. Mahajan,[1] H. Pirie,[1]

Z. Ahmed,[1] G. Bian,[3] X. Wang,[4,5] T.-C. Chiang,[5] and S. A. Brown[1, *]

[1] *The MacDiarmid Institute for Advanced Materials and Nanotechnology,*
*Department of Physics and Astronomy, University of Canterbury,*
*Private Bag 4800, Christchurch 8140, New Zealand*

[2] *Department of Solid State Physics, Faculty of Physics and Applied Informatics,*
*University of Lodz, 90-236 Lodz, Pomorska 149/153, Poland*

[3] *Department of Physics and Astronomy, University of Missouri, Columbia, Missouri 65211, USA*

[4] *College of Science, Nanjing University of Science and Technology, Nanjing 210094, China*

[5] *Department of Physics, University of Illinois at Urbana-Champaign,*
*1110 West Green Street, Urbana, Illinois 61801-3080, USA*

(Dated: August 14, 2017)

| | lattice constants | | | | | puckering | |
|---|---|---|---|---|---|---|---|
| | experimental | moiré sim. | calc. | lit. | bulk | calc. | lit. |
| 1 ML $\beta$-Sb | 404 ± 4 | 413 | 400 | 394 - 413 [1–6] | 431 Sb (111) [11] | 162 | 155-167 [1, 3–5] |
| 2 ML $\alpha$-Sb | 429 ± 7 × 486 ± 11 | 426×487 | 428×464 | ∼430×475 [2, 5] | 431×451 Sb (110) [11] | ≤ 20 | - |
| 2 ML $\alpha$-Bi | 453 ± 5 × 487 ± 4 | 453×487 | - | ∼450×480 [7] 440×478 [8] | 454×475 Bi (110) [12] | - | ≤5 [8]-60 [9, 10] |

TABLE SI. Lattice constants and puckering of the materials in this work. All values are given in picometers. We present our experimentally determined lattice constants (from Fourier analysis of atomically resolved topography images), the values used in the moiré simulations in Fig. 2 of the main report, as well as those obtained by our first principles calculations. In the hexagonal 1 ML-$\beta$-Sb structure, the puckering describes the distance between the two atomic planes; for the $\alpha$-structures it refers to the difference in the vertical position as defined in Ref.[8] In addition, we give values from the literature for the lattice constants and puckering of the thin films, and lattice constants for the respective planes in the bulk materials.

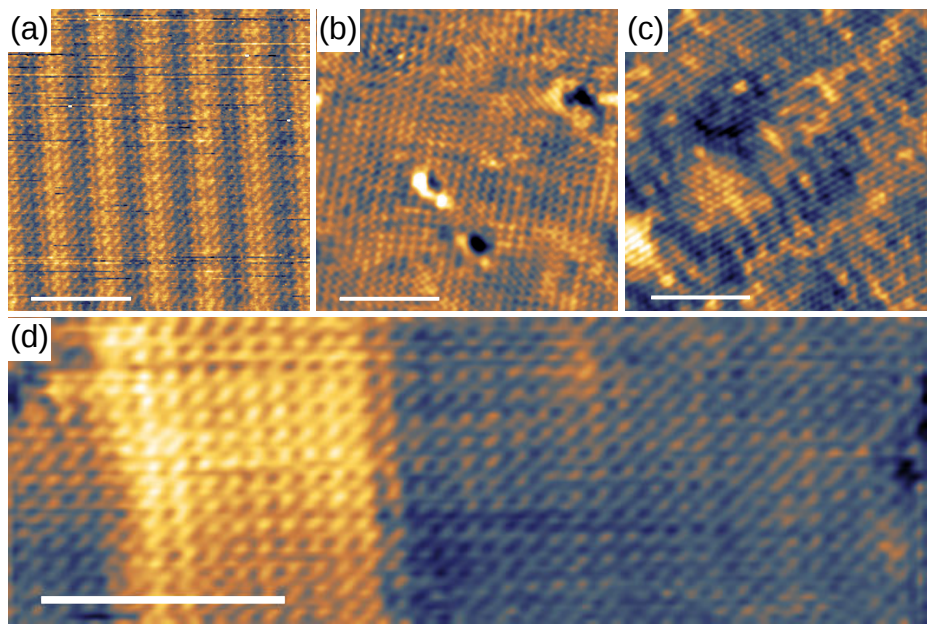

Fig. S1. Atomic resolution images of **(a)** the $\alpha$-Bi base, **(b)** 2 ML-$\alpha$-Sb, and **(c)** 1 ML-$\beta$-Sb . These images show the same regions as in Figs. 1(e)-(g) in the main paper, but on a larger scale so that the moiré patterns are more easily visible in (a) and (c). **(d)** close-up of the region marked by a blue rectangle in Fig. 1(a), showing atomic resolution on the narrow $\alpha$-Bi stripe (bright yellow) and the surrounding 2 ML-$\alpha$-Sb region (blue-orange). Clearly, the lattices of the two $\alpha$-phases are aligned. Raw STM images were filtered to enhance the contrast, scale bars: 5 nm.

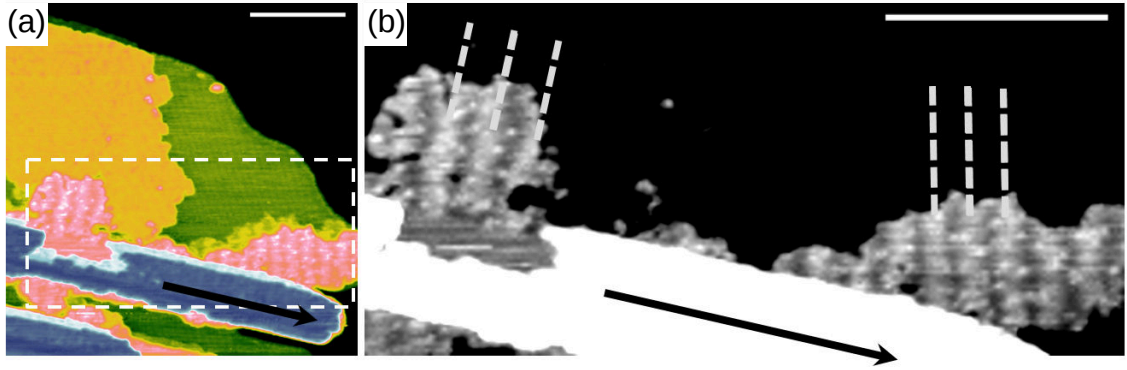

Fig. S2. Moiré patterns of 2 ML-$\alpha$-Sb on Bi(110) grown on HOPG showing that very similar moiré patterns are also observed for heterostructures grown on other substrates, and showing that the moiré patterns in the two different parts of the 2 ML-$\alpha$-Sb are at different angles due to a small rotation of the $\alpha$-Sb (see Figure S3). **(a)** STM topography image of an island grown on HOPG (black), where 1 ML-$\beta$-Sb (orange) and 2 ML-$\alpha$-Sb (pink-white) coexist on top of a bismuthene base (green). Note that the 5 ML-Bi stripe is not directly visible as it is covered by 2 ML-$\alpha$-Sb (blue). However, a faintly visible narrow 7 ML-Bi stripe (dark blue) is present, which allows determination of the preferred $\langle 1\bar{1}0 \rangle$ direction (black arrow). **(b)** Cropped STM image from (a), where the moiré patterns are enhanced (z-scale: 2 Å). Dashed lines indicate the orientation of the moiré patterns in two different 2 ML-$\alpha$-Sb regions. Imaging conditions: $+200$ mV, 10 pA, 295 K. Scale bar: 50 nm.

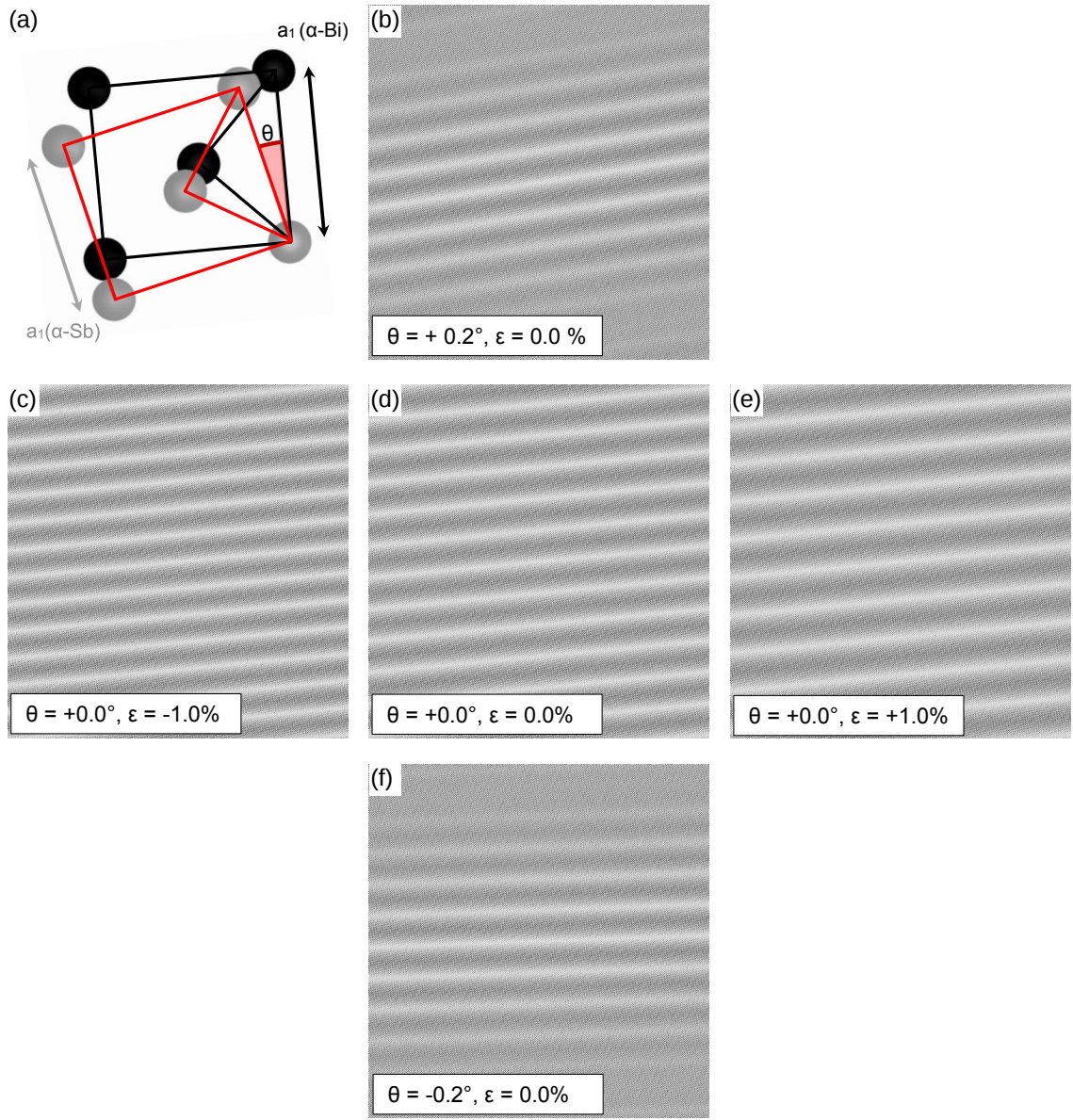

Fig. S3. Moiré sensitivity for the case of $\alpha$-Sb on $\alpha$-Bi: The model in **(a)** shows the $\alpha$-Bi unit cell (black) $\alpha$-Sb unit cell (grey, red) and the rotation angle $\theta$ of the overlayer; simulated moiré patterns are presented in panels **(b)** to **(f)** (size $77.6\,\mathrm{nm} \times 80.9\,\mathrm{nm}$). **(d)** shows the moiré pattern as in Fig. 2 and Fig. S4, using $426\,\mathrm{pm} \times 487\,\mathrm{pm}$ for 2 ML-$\alpha$-Sb, and $453\,\mathrm{pm} \times 487\,\mathrm{pm}$ for $\alpha$-Bi. Moiré patterns that result when $a_1^{\alpha\text{-}Sb}$ ($426\,\mathrm{pm}$) is strained by $\epsilon = \pm 1\,\%$ are shown in **(c)** and **(e)**, respectively. The effect of a relative rotation of $\theta = \pm 0.2°$ is shown in panels **(b)** and **(f)**. Similar sensitivity is observed for $\beta$-Sb on $\alpha$-Bi (not shown).

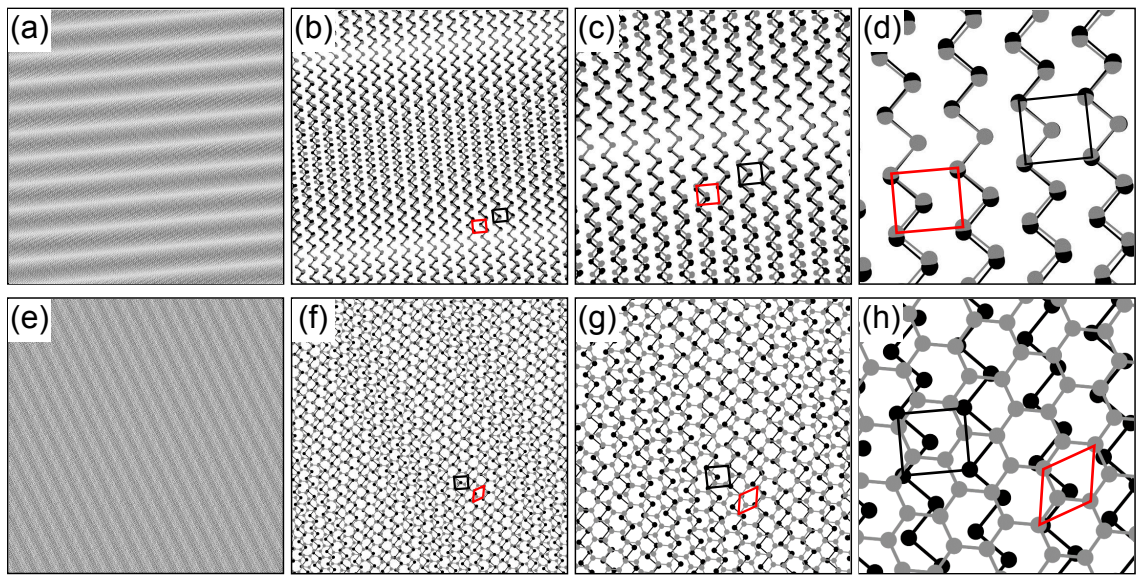

Fig. S4.  Simulated moiré patterns with ball-and-stick models for Sb (grey) on $\alpha$-Bi (black), showing the alignment of the two atomic crystals on different scales. **(a-d)** show 2 ML-$\alpha$-Sb on $\alpha$-Bi , (a) is the same as the simulated pattern in Figs. 2(a) and S3(d); **(e-h)** show 1 ML-$\beta$-Sb on $\alpha$-Bi. Unit cells are indicated in red for Sb and black for Bi; the width of the panels is 77.6 nm (a/e), 10 nm (b/f), 6 nm (c/g), and 2 nm (d/h). Lattice constants are 453 pm×487 pm for $\alpha$-Bi, 426 pm×487 pm for 2 ML-$\alpha$-Sb, and 413 pm for 1 ML-$\beta$-Sb (as in Table SI).

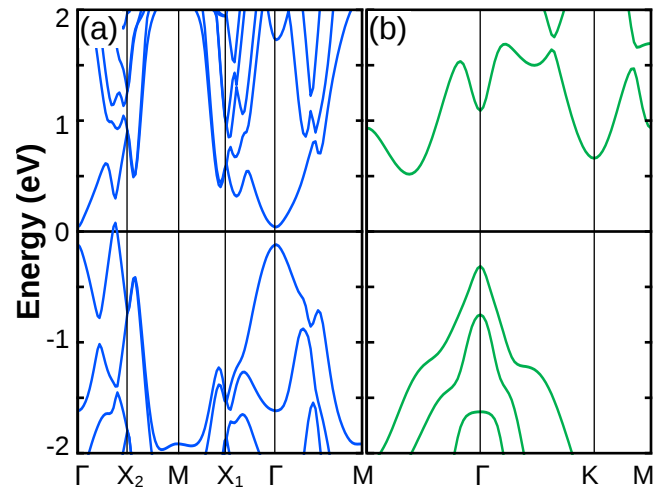

Fig. S5.   Electronic structure. Calculated band structures of **(a)** 2 ML-$\alpha$-Sb and **(b)** 1 ML-$\beta$-Sb along the indicated high-symmetry directions. The Fermi energy corresponds to zero in the diagrams.

| 2 ML-$\alpha$-Sb TRIMs | Parity eigenvalues | Parity product |
|---|---|---|
| Γ (0.0, 0.0, 0.0) | + − + − + − − + + − | − |
| X₂ (0.0, 0.5, 0.0) | − − − − − − − − − − | + |
| M (0.5, 0.5, 0.0) | + + − − − − − + + + | + |
| X₁ (0.5, 0.0, 0.0) | − − − − − − − − − − | + |
| 4 ML-$\alpha$-Sb TRIMs | Parity eigenvalues | Parity product |
| Γ (0.0, 0.0, 0.0) | + − + − + − + + − − − + − + − + − + + + | − |
| X₂ (0.0, 0.5, 0.0) | − − − − − − − − − − − − − − − − − − − − | + |
| M (0.5, 0.5, 0.0) | + + − − + + − − − − + + − − − + + + + − − | + |
| X₁ (0.5, 0.0, 0.0) | − − − − − − − − − − − − − − − − − − − − | + |

TABLE SII. Parity eigenvalues of valence bands of 2 ML and 4 ML $\alpha$-Sb films at time-reversal invariant momentum (TRIM) points.

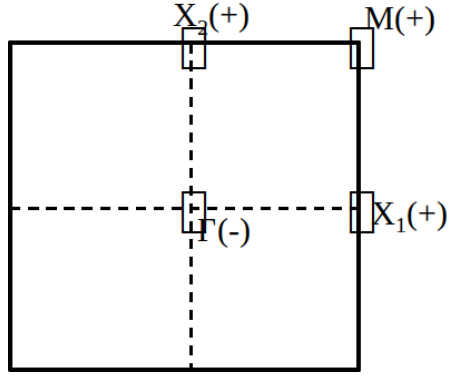

Fig. S6. The surface Brillouin zone of 2 ML and 4 ML $\alpha$-Sb films. The parity product at each TRIM point is indicated.

For centrosymmetric materials, the topological $\mathbb{Z}_2$ invariant can be calculated straightforwardly by examining the parity eigenvalues of the valence band at time-reversal invariant momentum (TRIM) points. According to Fu-Kane theory[13], the topological invariant is given by

$$(-1)^v = \prod_{i=1}^{TRIM} \delta_i \quad ,$$

where $\delta_i$ is the parity eigenvalue of the valence band at the TRIM point $\mathbf{k}_i$. We find that the $\mathbb{Z}_2$ invariant is $v = -1$ for both 2 ML and 4 ML $\alpha$-Sb films, identical to that of 2D topological insulators such as bismuthene, indicating that both 2 ML and 4 ML $\alpha$-Sb films have a topologically nontrivial band structure.

\* Corresponding author: simon.brown@canterbury.ac.nz



\end{document}